\documentclass[
 aip,
 amsmath,amssymb,
 reprint,%
]{revtex4-1}

\usepackage{graphicx}
\usepackage{dcolumn}
\usepackage{bm}
\usepackage{graphicx}
\usepackage{dcolumn}
\usepackage{bm}
\usepackage{physics}
\usepackage{amsmath}
\usepackage{amssymb}
\usepackage{csquotes}
\usepackage{array}
\usepackage{relsize}
\usepackage[utf8]{inputenc}
\usepackage[T1]{fontenc}
\usepackage{mathptmx}
\usepackage{etoolbox}
\usepackage{xcolor}
\usepackage{soul}
\usepackage[normalem]{ulem}
\newcommand{\stkout}[1]{\ifmmode\text{\sout{\ensuremath{#1}}}\else\sout{#1}\fi} 
\newcommand{\bt}[1]{\textcolor{blue}{#1}} 
\makeatletter
\def\@email#1#2{%
 \endgroup
 \patchcmd{\titleblock@produce}
  {\frontmatter@RRAPformat}
  {\frontmatter@RRAPformat{\produce@RRAP{*#1\href{mailto:#2}{#2}}}\frontmatter@RRAPformat}
  {}{}
}%
\makeatother
\begin{document}

\preprint{AIP/123-QED}

\title[]{ Coherent Nonlinear Optical Response for High-Intensity Excitation}
\author{Rishabh Tripathi}
\author{Krishna K. Maurya}
\author{Pradeep Kumar}
\author{Bhaskar De}
\author{Rohan Singh}
\email{rohan@iiserb.ac.in.}
\affiliation{ 
Department of Physics, Indian Institute of Science Education and Research Bhopal, Bhopal 462066, India.
}

\date{\today}

\begin{abstract}
Calculation of the coherent nonlinear response of a system is essential to correctly interpret results from advanced techniques such as two-dimensional coherent spectroscopy (2DCS).
Usually, even for the simplest systems, such calculations are either performed for low-intensity excitations where perturbative methods are valid and/or by assuming a simplified pulse envelope, such as a $\delta$-function in time.
Here, we use the phase-cycling method for exact calculation of the nonlinear response without making the aforementioned approximations even for high-intensity excitation.
We compare the simulation results to several experimental observations to prove the validity of these calculations.
The saturation of the photon-echo signal from excitons in a semiconductor quantum well sample is measured.
The excitation-intensity dependent measurement shows nonlinear contributions up to twelfth order.
Intensity-dependent simulations reproduce this effect without explicitly considering higher-order interactions.
Additionally, we present simulation results that replicate previously-reported experiments with high-intensity excitation of semiconductor quantum dots.
By accurately reproducing a variety of phenomena such as higher-order contributions, switching of coherent signal, and changes in photon-echo transients, we prove the efficacy of the phase-cycling method to calculate the coherent nonlinear signal for high-intensity excitation.
This method would be particularly useful for systems with multiple, well-separated peaks and/or large inhomogeneity.
\end{abstract}

\maketitle

\section{\label{sec:level1}Introduction}
Nonlinear coherent spectroscopy is a widely used technique for probing complex light-matter interactions at the quantum level \cite{Mukamel1995,Cundiff2008}.
A fundamental component of nonlinear spectroscopy is the wave-mixing process \cite{Shen2003,Bokor1981}, where a sequence of laser pulses interacts with the material to generate nonlinear signals based on the relevant nonlinear susceptibility of the material.
Among these interactions, four-wave mixing (FWM) is one of the most common, arising from the third-order susceptibility $\chi^{(3)}$ \cite{Vaughan2007}. 
Higher-order susceptibilities $\chi^{(n)}$ lead to wave-mixing processes such as six-wave mixing (SWM) and beyond \cite{Zhang2013, Gibson1991, Trebino1987, Tominaga1996, Steffen1996, Lucht1992, Kang2004}, which have been used to probe the electronic \cite{Oliver2015}  and vibrational states \cite{Bhattacharyya2019,Liu2019}.

Building on the wave\bt{-}mixing principle, multidimensional coherent spectroscopy (MDCS) extends this approach by measuring the wave-mixing signal field while varying two or more  time delays between interacting excitation pulses \cite{Li2023,Oliver2018,Hamm2011,GallagherFaeder1999,Cho2008,Brixner2017,Zhang2023,Zhang2013,Zhang2012a}. 
By performing Fourier transforms along the axes of these time delays, MDCS disentangles the nonlinear interactions of the material across multiple frequency dimensions. 
This multidimensional approach offers several advantages: it measures spectrally-resolved homogeneous response, isolates individual resonances in congested spectrum, and reveals the dynamic evolution trajectories of states that would otherwise remain hidden in one-dimensional measurements.
As a result, MDCS provides in-depth insights into phenomena such as energy transfer \cite{Thyrhaug2016,Bixner2012,Bressan2019}, chemical reactions \cite{Nuernberger2015}, and quantum coherence processes \cite{Butkus2017, Turner2010}.

The spectra obtained from MDCS are rich in information, revealing intricate features like coherent couplings \cite{Singh2014,Li2013a,Wen2013} and many-body interactions \cite{Shacklette2003,Erementchouk2007}. Proper interpretation of these complex signals requires robust simulations.
Conventionally,  theoretical and numerical models for simulation rely on a perturbative framework, where the nonlinear response scales linearly with the electric field of each excitation pulse, for weak excitation. 
In this framework, Double-sided Feynman diagrams (DSFDs) are often employed to map out specific quantum pathways \cite{Yang2007,Rose2021,Tekavec2007}, while response functions integrated with electric fields are used to compute the nonlinear response \cite{Mukamel1995,Gelin2013}.
However, the overall response is ultimately truncated to a specific order, as it assumes the perturbative expansion remains valid \cite{Gelin2009,Gelin2003,Suess2019}.
Consequently, MDCS experiments are generally conducted using low intensity laser pulses, ensuring that higher-order contributions remain below the noise level of measurements, keeping the analysis within the bounds of perturbation theory. 

On the other hand, high-intensity excitation can significantly influence coherent excitation and population transfer dynamics, as higher-order NWM processes become significant. 
MDCS experiments have demonstrated that intense laser field reveals various interesting phenomena across different systems, such as correlations among multiple particles \cite{Liang2021}, coherent control schemes through strong-field coupling \cite{Suzuki2016,Wen2013,Cerullo1996, Schneider2011}, coherence between highly excited states in quantum wells\cite{Turner2010}, and exciton-exciton annihilation (EEA) processes in excitonic systems \cite{Maly2018,Trinkunas2001}. 
Most theoretical descriptions of these phenomena rely on simplified approximations such as approximate electric-field envelopes of the excitation pulses \cite{Chen2017}, truncation of the signal up to a particular nonlinear order \cite{Hahn1998,Bressan2021,Binz2020,Anda2021} and/or neglecting pulse overlap effects \cite{Rao2017,Gelin2011}, which may not be valid for all cases.
Moreover, several studies have highlighted that finite pulse durations and overlaps can affect the MDCS spectra within the perturbative regime \cite{Smallwood2017, Rose2021,Do2017,Hybl2001,Li2013,Hedse2023}, suggesting that such factors could have a significant impact in non-perturbative scenarios too.
Recently, the ability to isolate specific signals such as the third-order response even in the presence of higher-order contributions was demonstrated in transient absorption experiments \cite{Maly2023}.
This technique has also been extended to 2DCS experiments performed in the pump-probe geometry \cite{Luettig2023, Luettig2023a}. 
Collectively, these observations underscore the need for a comprehensive approach that can accurately simulate nonlinear responses in both perturbative and non-perturbative regimes without relying on simplifying assumptions.

In this work, we numerically calculate the exact nonlinear signal  for high-intensity excitation using the phase-cycling method \cite{Tian2003,Tan2008,Zhang2012}.
In this approach, we do not make assumptions about the pulse envelope or truncate the signal up to a specific order.
We illustrate the above capabilities through numerical solutions of the optical Bloch equations (OBEs) for a few-level quantum systems.
The calculations are repeated by changing the phase of the excitation pulses and summing them up with specific weights to isolate the desired nonlinear signal.
We demonstrate that this approach offers exact solutions for nonlinear signals across both perturbative and nonperturbative regimes without relying on approximations of parameters such as pulse envelopes, pulse ordering, quantum pathways, and order of nonlinear interaction.
We begin, in Sec. \ref{nwm}, with a generalized description of the phase-cycling method for calculating an arbitrary $N$-wave mixing (NWM) signal, which will be used in the subsequent sections.
In Sec. \ref{saturation}, we model the saturation of the signal amplitude as the intensity of excitation pulses in 2DCS experiment is increased beyond the $\chi^{(3)}$ limit.
We further demonstrate the usefulness of this technique by reproducing previously reported experimental findings in quantum dots through phase cycling simulations in Sec. \ref{previous}.
Finally, we summarize our results in Sec. \ref{conclusion} along with a discussion of the significance of this work.
We note that methods similar to phase-cycling have been previously used to perform nonperturbative calculations of nonlinear optical response \cite{Gelin2022}, which, to the best of our knowledge, were limited to pump-probe signals \cite{Gelin2013,Gelin2011,Wang2008,Seidner1995}.

\section{N-wave-Mixing signal}
\label{nwm}

Nonlinear interaction of a system with a series of optical pulses results in a variety of nonlinear and linear responses.
Thus, for a meaningful interpretation of the results, it is critical to isolate a particular nonlinear signal.
Several techniques have been developed in order to achieve this goal both in experiments \cite{Hybl2001,Tian2003,Keusters1999,Gallagher1998} and simulations \cite{Gelin2009,GallagherFaeder1999,Mukamel2004}.
These techniques rely on some form of phase- and/or momentum-matching condition \cite{Bristow2009,Tekavec2007} and represent a general class of wave-mixing techniques; FWM is a well-known example.
In this section, we build on the phase-cycling method of calculating FWM signals to present a generalized formalism to calculate an arbitrary NWM signal.
While the discussion assumes that the light-matter interactions are perturbative in nature, in the subsequent sections we will demonstrate that the technique works even in highly nonpreturbative regime.

The signal in a NWM experiment is detected through either the $(N-1)^{th}$-order polarization or the the $N^{th}$-order population \cite{Nardin2015}.
For example, in a typical FWM experiment, the signal is radiated by the third-order polarization generated by three excitation pulses interacting with the sample.
This radiated field is interfered with a local oscillator to detect the phase-resolved signal.
Alternatively, a fourth excitation pulse can convert the third-order coherence to a fourth-order population and the desired signal is detected through frequency-domain filtering \cite{Tekavec2007} or phase-cycling \cite{Aeschlimann2011}.
The signal obtained for the two detection schemes may not be identical -- the population-detected signal may not provide phase information \cite{Gelin2022}, there might be quantitative differences due to ensemble averaging \cite{Anda2021}, and qualitative differences may appear due to the presence of dark states \cite{Smallwood2021}.
However, these differences are not relevant to the current study; thus, we consider the two detection schemes to be equivalent.

\begin{figure}[t]
    \centering 
    \includegraphics{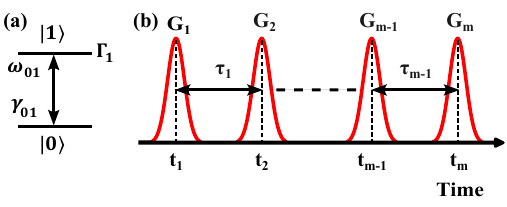}
    \caption{(a) Illustration of a two-level system showing the ground state $\ket{0}$ and the excited state $\ket{1}$ with a resonant frequency $\omega_{01}$.
    The coherence decay $\gamma_{01}$ and population decay $\Gamma_1$ terms are also indicated.
    (b) Pulse sequence showing $M$ excitation pulses.
    The delay between $i^{th}$ and $(i+1)^{th}$ pulses is $\tau_i$; unless specified, we use this convention throughout this work.}
    \label{fig:pulse}
\end{figure}

Phase-cycling methods and other simulation approaches \cite{Keusters1999,Gelin2022,Seidner1995} have previously been formulated for a general class of Hamiltonians.
However, for the purposes of this work, we have developed our formalism specifically for a two-level system, as it is most relevant to the systems studied here.
This formalism is extended later in the study to few-level systems. We consider a model two-level system (2LS), in Fig. \ref{fig:pulse}(a), which is excited by a sequence of $M$ pulses,  shown in Fig. \ref{fig:pulse}(b).
The $j^{th}$ pulse arrives at time $t=t_j$ with time- and position-dependent electric field
\begin{equation}
\label{eq:ej}
    E_j(t,\vb{r}) = \frac{1}{2}G_j(t-t_j)\left[ \exp{-i(\omega_j t -\vb{k}_j \cdot \vb{r} + \phi_{jM}} + c.c. \right],
\end{equation} where $G_j$ is the electric field envelope, $\omega_j$ is the frequency, $\vb{k}_j$ is the wavevector, $\phi_{jM}$ is the phase (defined with respect to the last pulse), and $c.c.$ denotes complex conjugate.
We consider degenerate excitation pulses and take $\omega_j = \omega_L$.
The total electric field incident on the sample is
\begin{equation}
\label{eq:etot}
    E_{tot}(t, \vb{r}) = \sum_{j=1}^M E_j(t, \vb{r}).
\end{equation}
The light-matter interaction in a 2LS can be described by the optical Bloch equations (OBEs):
\begin{subequations}
\label{eq:obes}
    \begin{align}
        {\dot{\rho}}_{11} &= -\Gamma_1 \rho_{11} - \frac{i\mu E_{tot}(t)}{\hbar} (\rho_{01}- \rho^*_{01}) \\
        {\dot{\rho}}_{01} &= -(i {\omega_{01}} + \gamma_{01})\rho_{01} + \frac{i\mu E_{tot}(t)}{\hbar} (\rho_{00}- \rho_{11}),
    \end{align}
\end{subequations}where $\rho_{00}$ and $\rho_{11}$ are density matrix elements representing the population in the ground $\ket{0}$ and excited $\ket{1}$ states, respectively.
$\rho_{01}$ and it{'}s complex conjugate $\rho^*_{01}$ represent the coherence between these states.
$\omega_{01}$ and $\mu$ are the transition frequency and dipole moment, respectively.
$\Gamma_1$ and $\gamma_{01}$ are population and coherence decay rates, respectively.
The population terms also satisfy an additional constraint for a closed system: $\rho_{00} + \rho_{11} = 1$.
Consistent with collinear geometry, we consider all the pulses to be propagating in the same direction; thus, the $\vb{k}$, and $\vb{r}$ terms have been dropped.
We numerically solve the OBEs under the rotating-wave approximation (RWA) \cite{Boyd2008} to obtain the exact state of the 2LS after excitation with the pulse sequence; additional details are provided in the Appendix \ref{app:Numerical_Solution}.
However, for meaningful interpretation of NWM experiment, a specific signal needs to be isolated.

A perturbative NWM signal can be seen as arising from the interaction of a sequence of optical pulses, where each pulse contributes to the signal through a specific interaction order.
The amplitude of the population-detected NWM signal
\begin{equation}
\label{eq:amp}
    S_{NWM} \propto \prod_{j=1}^{M} |E_{j}|^{|\alpha_j|},
\end{equation}
where $|E_{j}|$ represents the electric field amplitude of the $j^{th}$ pulse, and $|\alpha_j|$ indicates the order of interaction with the $j^{th}$ pulse.
The total number of interactions leading to the NWM signal is given by:
\begin{equation}
\label{eq:order}
    \sum_{j=1}^{M} |\alpha_j| = N
\end{equation}

The constants $\alpha_j$ can be either positive or negative and a collection of these define a specific $N$-wave mixing signal.
For example, a phase-matching condition within the framework of the RWA \cite{Berman2011,Mukamel1995}
\begin{equation}
\label{eq:pm}
    \phi_{sig} = \sum_{j=1}^{M} \alpha_{j}\phi_{jM},
\end{equation} defines a specific NWM signal. 

From the constraints outlined above, it is clear that in order to efficiently set up experimental conditions or perform simulations for isolating a specific NWM signal, it is essential to determine the minimum number of pulses required.
The minimum number of pulses is determined by the number of non-zero $\alpha_j$ coefficients in the phase-matching condition. For instance, in the case of a unique NWM signal where each pulse contributes to a single interaction, a minimum of $M = N$ excitation pulses are necessary to isolate the signal.
In cases where pulses interact more than once, the required number of pulses $M$ can be reduced.

Energy and momentum conservation are ensured through an additional constraint on the parameters $\alpha_j$:
\begin{equation}
\label{eq:constraint}
    \sum_{j=1}^{M} \alpha_{j} = 0.
\end{equation}
Consistent with the usual description of the collinear excitation scheme, the last pulse acts as a read-out pulse, which means
\begin{equation}
\label{eq:readout}
    \alpha_M = -1.
\end{equation}
Thus, we can rewrite the constraint in Eq. \eqref{eq:constraint} as $\sum_{j=1}^{M-1} \alpha_{j} = 1$.
This alternate framing explicitly indicates that after $M-1$ pulses, the signal corresponds to a polarization-detected signal of $(N-1)^{th}$ order, and signifies the equivalence of the population-detected signal and the heterodyne-detected radiated-polarization signal \cite{Gelin2022}.
Consequently, the phase matching condition defined in Eq. \eqref{eq:pm} is equivalent to signal emitted in the direction $\vb{k}_{sig} = \sum_{j=1}^{M-1}\alpha_j \vb{k}_j$ for noncollinear excitation schemes.

We will next outline the phase-cycling method to calculate the NWM signal $S_{NWM}$, which is essentially equal to the population $\rho_{11}$ obtained while satisfying the phase-matching signal in Eq. \eqref{eq:pm}.
The phase $\phi_{jM}$ is varied in equal steps within the range $[0, 2\pi)$.
For an arbitrary NWM signal, $\alpha_j$ can take $N+1$ possible values in the range $\left[-N/2, N/2\right]$.
Thus, $N+1$ phase-cycling steps for each excitation pulse is sufficient, although not necessarily optimal, to isolate any specific signal of interest.
However, for a generalized description, we consider that the phase of the $j^{th}$ excitation pulse is cycled in $P_j$ steps with the total number of phase-cycling steps
\begin{equation}
\label{eq:pcstep}
    N_{PC} = \prod_{j=1}^{M-1} P_j.
\end{equation}
The phase of the read-out pulse is fixed as 0, i.e. $P_M=1$.
A particular phase-cycling step $PC_r$ is defined by the collection of individual phases of the excitation pulses $$\left( \frac{2\pi p_1}{P_1}, \frac{2\pi p_2}{P_2},\ldots ,\frac{2\pi p_{M-1}}{P_{M-1}} \right)$$ where integer $p_j \in [0, P_j)$.
For each phase-cycling step, we solve the OBEs in Eqs. \eqref{eq:obes} to obtain $\rho_{11}(\tau_1, \tau_2, \ldots, \tau_{M-1}, r)$ where the integer $r$ represents the collection of phase terms of the excitation pulses.
The NWM signal is obtained through a weighted sum of the above solutions:
\begin{equation}
\label{eq:nwm}
    S_{NWM}(\tau_1, \tau_2, \ldots, \tau_{M-1}) = \sum_{r=1}^{N_{PC}} W_r \: \rho_{11}(\tau_1, \tau_2, \ldots, \tau_{M-1}, r)
\end{equation} with weights $$W_r = \prod_{j=1}^{M-1} \exp{-i\alpha_j \phi_{jM}(r)}$$ and phase of the $j^{th}$ pulse in the $r^{th}$ phase-cycling step is denoted by $\phi_{jM}(r)$.

In the above discussion we have assumed that the phase-cycling steps for each excitation pulse $P_j$ is known.
However, the choice of the optimal phase-cycling scheme with minimum $N_{PC}$ is not obvious and needs to be determined.
Such a phase-cycling scheme must ensure that aliasing of other signal does not happen \cite{Tan2008}.
We outline a generalized algorithm to perform the above check in Appendix \ref{app:PCM_valid}.
This aliasing check can be performed as the values of $P_j$ are varied to obtain the optimal phase-cycling scheme computationally.

We have made an implicit assumption in obtaining the NWM signal -- we have assumed that our 2LS is homogeneously broadened, i.e. there is a single resonance energy.
However, real systems often have inhomogeneity.
Let's assume that the inhomogeneous distribution is defined by function $q(\omega_m)$.
The NWM signal for this inhomogeneous system can be written as
\begin{equation}
\label{eq:inhom}
\begin{split}
    S_{NWM,in}&(\tau_1, \tau_2, \ldots, \tau_{M-1}) =\\
    &\sum_m q(\omega_m)S_{NWM}(\tau_1, \tau_2, \ldots, \tau_{M-1}, \omega_m),
\end{split}
\end{equation}where we have added explicit dependence of the homogeneous signal $S_{NWM}$ on the resonance energy $\omega_m$ \cite{Kumar2024}.
We will present results using this method when discussing the photon-echo signal in Sec. \ref{Inhomo_Photon echo}.

We can also extend the above procedure to a system with more energy states.
In this case we need to modify the OBEs in Eqs. \eqref{eq:obes} to appropriately describe the light-matter interactions.
Once a solution to these equations is obtained, we obtain the NWM signal by replacing $\rho_{11}$ in Eq. \eqref{eq:nwm} with a sum over the population in all the excited states.
This procedure will also be demonstrated later on, in Sec. \ref{Biexction} to simulate the coherent response of excitonic system including biexcitons.

Finally, we note that the procedure to calculate the NWM signal that we have discussed in this section did not make any approximations regarding the order of interaction, quantum pathways and pulse overlap.
Thus, we posit that the phase-cycling procedure can be used to calculate the coherent nonlinear response of the system even in the nonperturbative regime, as long as the equations describing the light-matter interactions and the RWA are valid.
In this respect, this procedure should directly mimic experimental condition, where the nonlinear signal in a particular direction or following a particular phase-matching condition can be detected even when the experiments are performed with intense excitation pulses.
As a consequence, it should be possible to explain results from high-intensity experiments by simply increasing the intensity of excitation pulses in simulations.
We prove this claim in the following sections by reproducing results from various experiments performed on semiconductor nanostructures  with excitation beyond the $\chi^{(3)}$ regime.
We clarify that a particular NWM signal is technically defined only for the $(N-1)^{th}$ order perturbative nonlinear signal.
As the excitation-intensity is increased, the NWM signal may not be truly isolated due to the higher-order contributions.
For instance, higher-order interactions such as $5^{th}$ and $7^{th}$-order may also contribute to the nonlinear response detected while satisfying the phase-matching condition for a $3^{rd}$-order signal.
However, for the sake of brevity, in the rest of this paper we refer to the nonlinear signal satisfying a specific phase-matching condition as the NWM signal even for high-intensity excitation when the perturbative treatment of light-matter interaction is not valid; this convention is consistent with previous work \cite{Fras2016}.

\section{Nonlinear Signal saturation}
\label{saturation}

In this Section, we use the formalism presented in Sec. \ref{nwm} to reproduce the experimentally-measured dependence of FWM amplitude on high-intensity excitation pulses.
We perform two-dimensional coherent spectroscopy (2DCS) experiment to measure the excitation-density-dependent FWM signal from a GaAs QW-QD nanostructure \cite{Gammon1996, Moody2011}.
The GaAs interfacial quantum dots (IFQDs) sample consists of a 4.2 nm wide GaAs/Al\textsubscript{0.3}Ga\textsubscript{0.7}As quantum well (QW) with single monolayer (ML) interface fluctuations formed by brief interruptions during epitaxial growth.
These fluctuations cause significant confinement changes, splitting the optical spectrum into two peaks: higher-energy, weakly-localized QW excitons and lower-energy excitons localized at the ML fluctuation sites, which form the IFQDs.
Our measurements are focused specifically on the QW excitations.

\begin{figure}[t]
\includegraphics{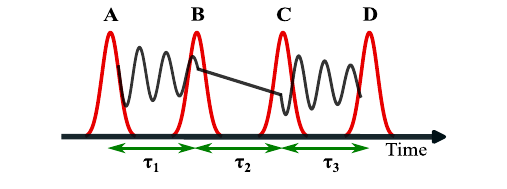}
\caption{The pulse sequence used in the experiment.
The black line illustrates the evolution of the system's density matrix elements: the first-order coherence induced by pulse A, second-order population created by pulse B, and third-order coherence generated by pulse C.
The fourth pulse D acts as a local oscillator (LO) to interfere with the radiated signal.
Delays between successive pulses are indicated by $\tau_1$, $\tau_2$, and $\tau_3$.
}
\label{fig:seq}
\end{figure}

\begin{figure}[t]
\includegraphics{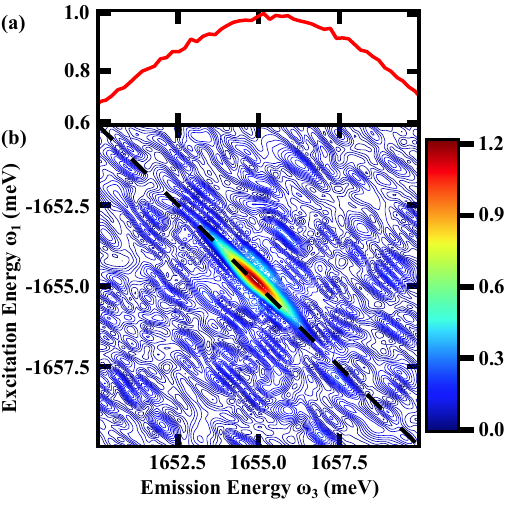}
\caption{(a) Normalized spectrum of the excitation pulses used in the 2DCS experiment.
(b) Absolute-value rephasing 2D spectrum for the GaAs QW-QD nanostructure, with each pulse intensity set at 0.5 $\mu$W.}
\label{fig:2d_spec}
\end{figure}

We use a collinear 2DCS experiment, which is similar to the commonly-used acousto-optic modulator (AOM) based experiments \cite{Martin2018,Nardin2013,Tekavec2007}, and is briefly described here.
The excitation pulses are derived from the output of an optical parametric amplifier (OPA) system with a repetition rate of 50 kHz.
The output of the OPA is centered at 755 nm and has a full-width at half-maximum (FWHM) bandwidth of $\sim 20$ nm, which results in a transform-limited pulse duration of $\sim 40$ fs.
These pulses are then split into four different paths (A, B, C and D), and are tagged with unique radio frequencies -- $\Omega_A=78.858$ MHz, $\Omega_B=78.595$ MHz, $\Omega_C=79.597$ MHz, and $\Omega_D=79.304$ MHz through phase modulation in the AOMs.
The drive frequencies for the AOMs are generated through a direct digital synthesizer and are delivered to the AOMs using high-voltage drivers.
These pulses are recombined to obtain the excitation pulses with a specific time-ordering that is shown in Fig. \ref{fig:seq}.
The delays between the consecutive pulses -- $\tau_1$, $\tau_2$, and $\tau_3$ can be scanned using motorized delay stages.
The FWHM bandwidth of excitation pulses is reduced to $\sim 6$ nm using a grating-based pulse shaper.
The narrow spectrum of the excitation pulse, shown in Fig. \ref{fig:2d_spec}(a), ensures that only the QW excitons are excited.
The excitation pulses with collinear polarization are focused onto the sample using a lens with a focal length of 20 cm, resulting in a spot size with a $1/e^2$ radius of $\sim100 \:\mu$m.
All the excitation pulses have equal intensity, which is controlled using a combination of a half-wave plate and a polarizer. 
A continuous wave (CW) laser with a wavelength of 1064 nm is used for passive stabilization of the pulse delays.
The CW laser traverses the same optical path as the excitation laser and is, thus, modulated at the same AOM frequencies.
Two photodiodes measure the beat notes arising from the interference of the CW laser from different paths -- $\Omega_{AB} = 263$ kHz from the A-B path and $\Omega_{CD} = 293$ kHz from the C-D path.
These beat notes provide the reference $\Omega_{FWM} = \Omega_{CD} - \Omega_{AB} = 30$ kHz to measure the rephasing FWM signal using a lock-in amplifier.
The FWM signal is measured as a function of variable delays $\tau_1$ and $\tau_3$ while keeping delay $\tau_2$ fixed at 0.7 ps.
Delay $\tau_1$ is scanned in discrete steps, while delay $\tau_3$ is scanned continuously to reduce the data-aquisition time.
The generation of the phase-stabilized reference, lock-in detection, and scanning of the time delays is performed using a customized Bigfoot electronics from MONSTR Sense that uses a field-programmable gate array (FPGA) \cite{Huang2023,Polczynska2023}.
The electronics system generates a virtual reference signal corresponding to a wavelength close to the exciton resonance, which reduces the phase fluctuations on the lock-in detected signal significantly.

Figure \ref{fig:2d_spec}(b) shows an absolute rephasing spectrum, with the excitation energy $\omega_1$ and the emission energy $\omega_3$ corresponding to a Fourier transform of the measured signal with respect to the time delays $\tau_1$ and $\tau_3$, respectively.
The absorption energy axis shows negative energy since the phase evolution during delay $\tau_1$ is opposite to that during time $\tau_3$ for the rephasing signal.
The spectrum features a diagonally elongated peak corresponding to an inhomogeneously broadened QW excitonic state, which is consistent with previous studies on the same sample \cite{Moody2011}.

\begin{figure}[b]
\includegraphics{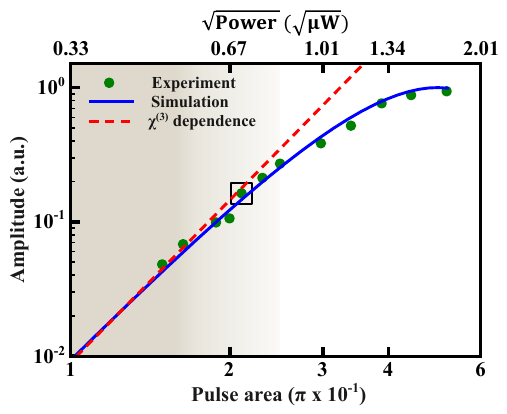}
\caption{Log-log plot of the amplitude of the experimental rephasing spectrum (green circles) as a function of the power of excitation pulses, and the amplitude of the simulated rephasing spectrum (solid blue line) versus the pulse area of the excitation pulses.
The data point encased in a box corresponds to the 2D spectra shown in Fig. \ref{fig:2d_spec}(b).
The dashed red line indicates the expected $\chi^{(3)}$ dependence. The shaded region marks the $\chi^{(3)}$ regime. 
The lower X-axis represents the pulse area for each point of the simulated 2D spectra, while the upper X-axis corresponds to the square-root of the power of the excitation pulses in the experiments.
The simulation data points are normalized by the peak amplitude at maximum pulse area and the experimental values  were rescaled  to match  the normalized simulation data.}
\label{fig:pwr}
\end{figure}

We measured rephasing spectra as a function of intensity of excitation pulses.
The intensity-dependent peak amplitude, which is calculated using the procedure outlined in Appendix \ref{app:noise}, is shown as a log-log plot in Fig. \ref{fig:pwr}.
Figure \ref{fig:pwr} also shows the expected $\chi^{(3)}$-dependence in dashed line.
At low intensities (shaded region in Fig. \ref{fig:pwr}), the signal amplitude follows this behavior, which is roughly in agreement with prediction of third-order perturbation theory.
However, as the average power per excitation pulse exceeds $\sim 0.7$  $\mu$W, the data deviates from the expected $\chi^{(3)}$-dependence and saturates.
This behavior suggests that at high intensities, third-order perturbation theory does not accurately describe the FWM signal amplitude, and higher-order nonlinear signals contribute significantly to the total signal.

We reproduce the experimental results using simulations based on the phase-cycling method discussed in Sec. \ref{nwm}.
We model the excitonic system as a two-level system, as shown in Fig. \ref{fig:pulse}(a).
We consider the sum of the electric fields of four excitation pulses, as shown in Fig. \ref{fig:seq}, and numerically solve the OBEs in Eqs. \ref{eq:obes} as a function of delays $\tau_1$ and $\tau_3$.
The most appropriate and economical $3 \times 3 \times 3 \times 1$ phase-cycling scheme was used to acquire the rephasing signal as a function of the variable delays.
The 2D rephasing spectrum was then obtained by performing a Fourier transform with respect to both the time delays.
Various rephasing spectra were simulated with increasing pulse areas of each excitation pulse.
The log-log plot of the maximum amplitude of each spectrum against the pulse area, indicated by the solid blue line in Fig. \ref{fig:pwr}, shows an excellent match with the data for $\mu \approx 630$ D, which is consistent with previously reported value for QW excitons \cite{Christmann2011}.

One can argue that it should be possible to reproduce the observed saturation behavior by using higher-order perturbation terms.
Indeed it is possible to reproduce the experimental finding using a polynomial fit that includes significant contribution from all the orders up to twelve-wave mixing signal.
Simulating this high-order response will be extremely complicated due to the large number of possible quantum pathways.
The relative strength of the different orders of interaction is also difficult to predict a priori.
Furthermore, the validity of applying a perturbative calculation to such high-order interactions is tenuous, at best.
The phase-cycling method provides an exact solution of the nonlinear signal and, thus, bypasses these issues while demonstrating that the saturation behavior is a natural consequence of increasing the intensity of excitation pulses.

\section{Phase-Cycling Simulations of Previous Experimental Works}
\label{previous}
In this section, we aim to further demonstrate the effectiveness of the phase-cycling method by reproducing several results from previously reported high-intensity coherent nonlinear spectroscopy studies performed on InAs self-assembled QDs.
We start each subsection with a brief description of the results that are relevant for this work.
We then reproduce these results without making approximations about the pulse envelope or truncating the signal to a specific perturbation order.
In addition to proving the validity of the phase-cycling method in the high-intensity regime, we also provide a pathway to correctly model the response of a system where the above approximations may not be valid.

\begin{figure}[b]
\includegraphics{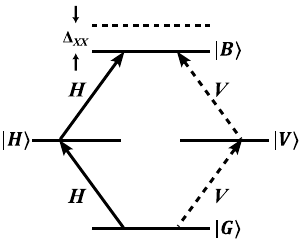}
\caption{An energy-level diagram for the exciton-biexciton diamond-type system depicting the ground state $\ket{G}$, the lowest-energy exciton states $\ket{H}$ and $\ket{V}$, and the bound biexciton state $\ket{B}$, which is red-shifted from the unbound two-exciton state by the biexciton binding energy $\Delta_{XX}$.
The $\ket{H}$ and $\ket{V}$ exciton states are excited by light with orthogonal linear polarizations H and V, which are indicated by the solid and dashed arrows, respectively.}
\label{fig:bx}
\end{figure}

\subsection{Higher-order features}
\label{Biexction}

We begin with a study \cite{Moody2013} where $\chi^{(5)}$ signal from the biexciton state in self-assembled InAs QDs was spectrally isolated in a noncollinear 2DCS experiment.
This study used cross-linearly polarized (HVVH) excitation sequences to selectively excite the biexcitonic state in neutral QDs.
The FWM signal was measured in the phase-matched direction $\vb{k}_S = -\vb{k}_A + \vb{k}_B + \vb{k}_C$.
At low excitation intensities, i.e. within the $\chi^{(3)}$ regime, the 2D spectrum displayed a single biexciton peak.
However, at higher intensities, an additional biexciton peak emerged, which was attributed to the $\chi^{(5)}$ nonlinear response.
DSFDs, with contributions up to fifth-order excitation were used to explain the origin of this peak.

\begin{figure}[b]
\includegraphics{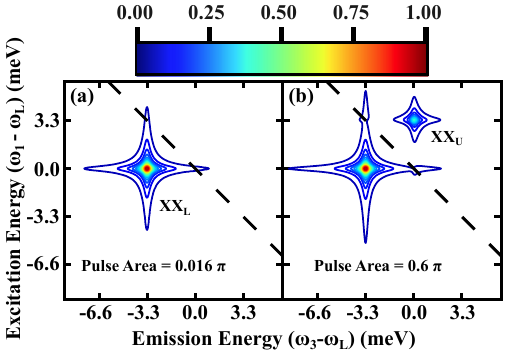}
\caption{ Normalized rephasing amplitude spectra for cross-linear (HVVH) polarization are shown for excitation pulse areas of (a) $0.016 \pi$ and (b) $0.6 \pi$.
The peak in (a) corresponds to the $\chi^{(3)}$ response involving the biexciton resonance.
The additional peak in (b) was attributed to the $\chi^{(5)}$ contributions in Ref. \onlinecite{Moody2013}. Both the excitation and emission axes are shown relative to the center frequency of the excitation pulse $\omega_L$.}
\label{fig:bx2d}
\end{figure}

\begin{figure}[t]
\includegraphics{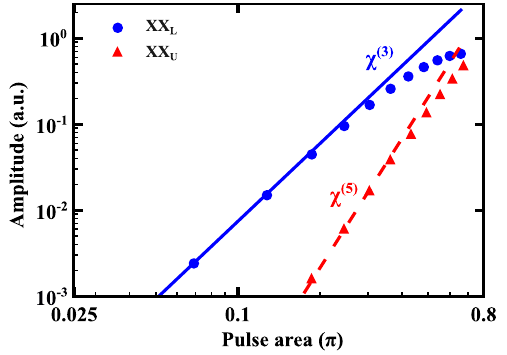}
\caption{Power dependence measurements for both biexciton peaks are shown, illustrating the amplitudes of the peaks versus the pulse area of the excitation pulses.
The $XX_L$ peak, represented by blue dots, initially shows $\chi^{(3)}$ dependence (blue solid line).
As this peak saturates, the $XX_U$ peak appears ( red triangles) and shows $\chi^{(5)}$ dependence ( red dashed line) at intermediate intensities.
Ultimately, the $XX_U$ peak also shows saturations at even higher intensity of excitation pulses.}
\label{fig:bx_pwr}
\end{figure}

We model the exciton-biexciton system as a diamond-type system shown in Fig. \ref{fig:bx}, and the OBEs were modified accordingly \cite{Suzuki2016}(see Appendix \ref{app:OBEs_diamond} for the detailed equations).
We incorporated polarization of excitation pulse in the electric fields used in these equations to ensure a cross-linear excitation scheme.
For simplicity, the trion states, inhomogeneity, and the fine-structure splitting were ignored since they are not relevant for the results discussed here.
Biexciton binding energy $\Delta_{XX} = 3.3$ meV was used.
Similar to the previous section, the $3 \times 3 \times 3 \times 1$ phase cycling method was chosen to simulate the rephasing signal satisfying the phase matching condition $\phi_{sig} = -\phi_A + \phi_B +\phi_C - \phi_D$.
We varied the pulse area of the excitation pulses in our simulation.
At low intensities, consistent with the $\chi^{(3)}$ regime, the simulation results showed a single biexciton peak XX\textsubscript{L}, as depicted in Fig. \ref{fig:bx2d}(a).
As the pulse intensity increased, the simulated spectrum in Fig. \ref{fig:bx2d}(b) revealed the emergence of an additional biexciton peak XX\textsubscript{U}.
These peaks are shifted from the diagonal along the excitation and emission axes by the biexciton binding energy, which is consistent with the experiment.
As was done in the original study, a comparison of the excitation-intensity dependence of the peak amplitudes highlights the distinct perturbative order contributions of $\chi^{(3)}$ and $\chi^{(5)}$ to these peaks, as shown in Fig. \ref{fig:bx_pwr}.

Our calculations reproduced the experimental results without truncating the calculations to a particular order of light-matter interaction.
Furthermore, the deviation of both peak amplitudes from the $\chi^{(3)}$ and $\chi^{(5)}$ lines suggests that higher-order non-linear contributions ($\geq \chi^{(7)}$ ) become significant at higher intensities; this behavior was observed in the experimental data, but could not be replicated with the perturbative theory used in the earlier work.

\subsection{Switching between nonlinear signals}
\label{FWM-SWM}

We next focus on a study \cite{Fras2016} that reported conversion of FWM signal to SWM signal in a single self-assembled InAs QD, showcasing a novel coherent-control mechanism in single quantum emitters.
The researchers employed three collinear excitation pulses, each uniquely tagged with a frequency $\Omega_i$ ($i \in {1,2,3}$) using AOMs, to resonantly drive a single InAs quantum dot.
The first two pulses create a FWM transient, which evolves freely until the arrival of the third pulse.
This third pulse converts the FWM signal $\xi_F$, detected at the heterodyne frequency $\Omega_F = 2\Omega_2 - \Omega_1$  into SWM signal $\xi_S$, detected at the heterodyne frequency $\Omega_S = \Omega_1 - 2\Omega_2 + 2\Omega_3$, achieving up to unity conversion efficiency.
This conversion process depends on the delay and pulse area of the third pulse, aligning with theoretical model derived by assuming excitation with $\delta-$function pulses:
\begin{subequations}
\label{eq:fwm-swm}
    \begin{align}
    \left| \xi_F(\tau_3)  \right| &= \sin{\theta_1} \sin^2\left({\frac{\theta_2}{2}}\right) \left[ \Theta(\tau_3 + \tau_2) - \sin^2\left({\frac{\theta_3}{2}}\right) \Theta(\tau_3) \right], \\
    \left| \xi_S(\tau_3)  \right| &= \sin{\theta_1} \sin^2\left(\frac{\theta_2}{2}\right) \sin^2\left(\frac{\theta_3}{2}\right) \Theta(\tau_3).
    \end{align}
\end{subequations}
In the above equations, $\theta_i$ is the pulse area of the $i^{th}$ excitation pulse and $\Theta$ is the Heaviside function.
According to these equations, when the pulse area of the third pulse $\theta_{3}$ is set to $\pi$, the FWM signal field $\xi_F(\tau_3)$ should become zero at $\tau_3 = 0$, while the SWM signal field $\xi_s(\tau_3)$ should reach its maximum value, which implies a complete conversion of the FWM signal to the SWM signal.

\begin{figure}[t]
\includegraphics{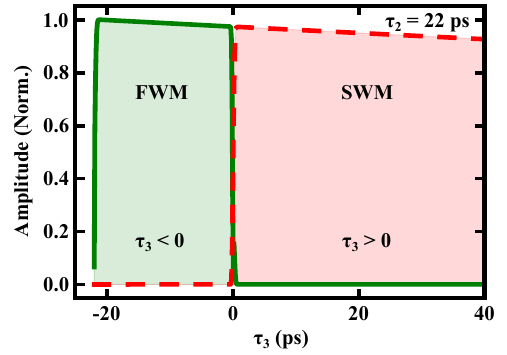}
\caption{Amplitude of FWM and SWM signals as a function of delay, $\tau_3$ (the delay between the third and fourth pulses). For negative delays ($\tau_3 < 0$), only the FWM signal is present, indicated by the green shaded area under the curve. For positive delays ($\tau_3 > 0$), the FWM signal fully converts to SWM, shown by the red shaded area under the curve, demonstrating complete FWM-to-SWM conversion.}
\label{fig:FWM_SWM swapping}
\end{figure}

To replicate these observations, we perform simulations with four excitation pulses and calculated the FWM and SWM signals with phase-matching conditions $\phi_F = -\phi_1 + 2\phi_2 -\phi_4$ and $\phi_S = \phi_1 -2\phi_2 + 2\phi_3-\phi_4$.
The signals were calculated as a function of the delay $\tau_3$ between the third excitation pulse and the read-out pulse while keeping delays $\tau_1$ and $\tau_2$ constant.
We vary delay $\tau_3$ from negative to positive values to replicate the experiment.
For negative values of $\tau_3$ the readout pulse arrives before the third excitation pulse and we can only obtain FWM signal.
As $\tau_3$ becomes positive, the sample interacts with the third-interaction pulse and it is possible to obtain SWM signal too.

The optimal phase cycling methods for the signals corresponding to $\phi_F$ and $\phi_S$ were $3\times6\times3\times 1$ and $3\times5\times4\times 1$, respectively.
However, we used the $3\times6\times4\times 1$ phase cycling method, which allowed us to isolate both signals simultaneously using the same phase combination of excitation pulses.
This phase-cycling scheme is the optimal choice for our purpose.
As an added advantage, using the same phase-cycling scheme for both the FWM and SWM signals allows a direct comparison of their amplitudes.
This approach is also relevant for experimental settings where the same phase combination can be used to simultaneously measure both signals.
We model a single QD as a 2LS and used the OBEs incorporating all decay terms similar to those in the discussed study.
Our simulation results, depicted in Fig. \ref{fig:FWM_SWM swapping} for $\theta_3 = \pi$, show similar observations to those described in the study.
During the time interval between the arrival of the second and third excitation pulses, the FWM signal decays with the population decay rate (0.01 meV) while the SWM signal amplitude remains zero.
Consistent with the analytical calculations, we observed that as the third pulse arrived ($\tau_3>0$ ps), the total FWM signal is converted to the SWM signal, and the FWM signal value dropped to zero.

The above result proves that we can accurately measure the response of the system even in highly nonperturbative regime using the phase-cycling scheme.
We also note that in the original work \cite{Fras2016}, the authors arrived at an analytical solution by assuming $\delta$-function excitation pulses in the time domain, which is significantly simpler than our approach.
However, this assumption may not be valid in case of significant inhomogeneity and/or well separated, multiple resonances. In such cases, the influence of pulse duration and overlap can no longer be ignored. The use of $\delta$-function pulses would not accurately replicate experimental results, and theoretical expressions for more realistic pulse envelopes like Gaussian; become impossible to calculate analytically. In contrast, 
the numerical calculations that we have presented provide a method to calculate the nonperturbative response without making the above assumption.

\subsection{Photon echo from high-intensity excitation}
\label{Inhomo_Photon echo}

As a final example, we look at a study that investigated the effect of high-intensity excitation pulses on the photon-echo transients in an ensemble of (In,Ga)As/GaAs QDs \cite{Poltavtsev2016}.
In a noncollinear two-pulse FWM experiment, the photon-echo signal peaks after the second excitation pulse at a delay equal to the delay between the excitation pulses.
In Ref. \onlinecite{Poltavtsev2016}, experiments performed with high-intensity excitation pulses show that the photon-echo transients do not necessarily follow this behavior.
Solutions of the Lindblad equation assuming rectangular temporal profile of the excitation pulses were used to model the experimental findings and infer that significant modifications in the photon-echo transients occur when the spectral width of the inhomogeneous ensemble is comparable to those of the excitation pulses. 

\begin{figure}[t]
\includegraphics{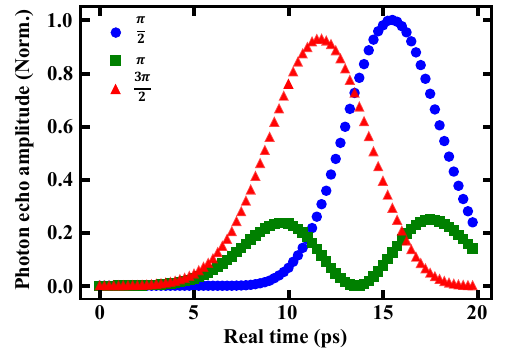}
\caption{Simulated transient photon echo for three different pulse areas of the first pulse: $\theta_1$ = $\pi /2$, $\pi$, and $3\pi/2$}
\label{fig:Inhomo_Photon echo}
\end{figure}

We were also able to reproduce these results through phase-cycling simulations.
The QDs are modeled as a 2LS and we use three excitation pulses to calculate the $2\vb{k_2}-\vb{k_1}$ signal.
We employed a $5\times2\times1$ phase cycling scheme to isolate the FWM signal with the phase matching condition $\phi_{sig} = -\phi_1 +2\phi_2-\phi_3$. In our simulation, we set the pulse duration $\tau_p = 2.9$ ps and the standard deviation of inhomogeneity $\delta=0.276$ meV, resulting in a product $\tau_p\delta$ = 1.2, consistent with the calculations in the original study.  The rephasing pulse had a pulse area of 
$\pi$, and the delay between the first two pulses was fixed at $\tau_1=15$ ps.
The absolute value of the transient FWM signal was measured as a function of the delay between the second and third pulses.
The simulation results are shown in Fig. \ref{fig:Inhomo_Photon echo} for various pulse areas of the first pulse $\theta_1 \in \left\{\frac{\pi}{2}, \pi, \frac{3\pi}{2}\right\}$.
These results are consistent with the previously reported experimental observations and calculations \cite{Poltavtsev2016}. For the $\frac{\pi}{2}$ pulse area, the photon echo appears slightly delayed relative to the expected time (15 ps), which equals the fixed delay $\tau_1$.
However, for the $\frac{3\pi}{2}$ pulse, the FWM signal shows an advanced photon echo (before 15 ps), while for $\theta_2 = \pi$ the PE transient is significantly diminished and comprises of two peaks.

The above results explicitly demonstrate that the our calculations also work for systems with inhomogeneous broadening.
As in Sec. \ref{FWM-SWM}, we did not make any simplifying assumptions about the pulse shape, as was done in the original work \cite{Poltavtsev2016}.
Collectively, the results presented in this section highlight the effectiveness of the phase-cycling method to calculate the exact nonlinear response even for high-intensity excitation pulses.
Interestingly, the calculations in in Secs. \ref{FWM-SWM} and \ref{Inhomo_Photon echo} were performed for excitation pulses with pulse area $\theta \geq \pi$.
At such high-intensities, perturbative calculations cannot be used even if perturbative contributions up to an arbitrarily high order are included.
The fact that we can reproduce experimental results in this regime further highlights the method's usefulness.
While the formalism presented here assumes collinear pulses and population-detected signals, the experimental results reproduced in this study were based on heterodyne-detected radiated-polarization signals. 
This underscores the broader applicability of our approach irrespective of the detection scheme of signal used in experiments.

\section{Conclusion}
\label{conclusion}
In this work we have presented a comprehensive study on the application of the phase-cycling method to simulate coherent nonlinear signal for high-intensity excitation pulses.
We started with a description of the phase-cycling method for a 2LS that can be applied to an arbitrary NWM signal.
Following this, we used the phase-cycling method to reproduce the experimentally-measured saturation of the FWM response of QW excitons as the excitation pulse intensity is increased.
Finally, we reproduced three previously-reported experimental results on measuring coherent nonlinear response of QDs using high-intensity excitation pulses, including experiments performed in nonperturbative regimes with pulse area of excitation pulses $\sim \pi$.

This study highlights the advanced capabilities of the phase-cycling method in providing exact solutions for nonlinear signal.
Unlike conventional approaches that often rely on simplifying assumptions, such as the use of Feynman diagrams to consider relevant quantum pathways, or assumptions about pulse shapes, the phase-cycling method handles all these complexities inherently.
Prior studies have have attempted to address limitations of the conventional perturbative framework by introducing finite pulse solutions to examine effects such as narrow spectral bandwidths \cite{Abramavicius2010,Leng2017,Schweigert2008}, chirp \cite{Smallwood2017,Binz2020}, and pulse overlap \cite{Hybl2001,Li2013,Hedse2023} on the total response.
However, these methods typically address a single factor at a time and still rely on truncating the response at a specific order. 
In contrast, the phase-cycling approach is effective across both perturbative and non-perturbative regimes, allowing us to simulate these pulse effects accurately without the need for such approximations.

In addition to the results discussed in this work, coherent interaction of high-intensity pulses with two- or few-level systems can be interesting to understand several other phenomena, such as many-body interactions \cite{Meinardi2003}, exciton-exciton annihilation \cite{Tiwari2012}, overcoming electronic dephasing and vibrational relaxation \cite{Egorova2008}, and manipulating nonadiabatic couplings \cite{Chen2022}.
Beyond these, V-type systems, which naturally occur in materials such as transition metal dichalcogenides (TMDCs) \cite{Wang2008}, semiconductor nanostructures \cite{Wang2005}, and atoms \cite{Lim2011}, also provide opportunities to investigate even richer dynamics.
These systems feature excited states coupled through a common ground state, enabling studies of quantum interference among multiple-order signals under high-intensity conditions \cite{Wu2024,Binz2020}.
This work can also be relevant to understand the effects of high-intensity exciation on complex phenomena such as electron-vibrational coupling in molecular systems where the Hamiltonian can be reduced to consider a few eigen states \cite{Lewis2012,Duan2017}.

Finally, we address some limitations of the simulations performed in this work, which are primarilty related to the validity of the OBEs and the RWA.
For instance, the OBEs are not valid in case of non-Markovian decay dynamics \cite{Berman1986} and also do not include multiphoton excitation terms that may be relevant for high-intensity excitation \cite{Kay1981}.
Similarly, the RWA will not be valid for high-intensity excitation as the Rabi frequency $\Omega_R$ approaches or exceeds the pulse carrier frequency $\omega_L$ \cite{Berman2011}.
Our simulations align well with the experimental results discussed, confirming the validity of the RWA under these conditions.
We also note that the phase-matching condition is a manifestation of the RWA; the wave-mixing experiments, as are commonly interpretated, would not work if the RWA is not valid.
Nevertheless, it is instructive to estimate the region of validity of the simulations from the perspective of the RWA.
While the Rabi frequency is typically defined for continuous-wave (CW) lasers, it is not straightforward to define for pulsed lasers \cite{Allen1987}.
However, analogous to CW excitation, we can define average Rabi frequency $\langle \Omega_R \rangle = \theta/\Delta t$ for pulsed excitation in terms of the pulse area $\theta$ and FWHM $\Delta t$.
Using this approximation, for an optical laser pulse centered at $\omega_L = 800$ nm with $\Delta t = 100$ fs and $\theta = 2\pi$, we obtain the ratio $\omega_L / \langle \Omega_R \rangle \approx 40$.
Thus, the RWA and, consequently, our calculations should hold even for the above situation, which assumes a significantly higher $\langle \Omega_R \rangle$ compared to those used in this work.

\appendix
\section{Numerical Simulation Framework}
\label{app:Numerical_Solution}
Our simulation framework begins with the reformulation of the OBEs in the rotating frame of reference, utilizing the RWA to retain only resonant terms, as exemplified for a two-level system by:
\begin{subequations}
\label{eq:obes_rot}
    \begin{align}
        {\dot{\rho}^{'}}_{11} &= -\Gamma_1 \rho^{'}_{11} - \frac{i\mu G_{tot}(t)}{2\hbar} (\rho^{'}_{01}- \rho^{'*}_{01}) \\
        {\dot{\rho}^{'}}_{01} &= -(i \Delta_{01} + \gamma_{01})\rho^{'}_{01} + \frac{i\mu G_{tot}(t)}{2\hbar} (\rho^{'}_{00}- \rho^{'}_{11}),
    \end{align}
\end{subequations} Here, $\rho^{'}_{ij}$ = $\rho_{ij} e^{-i\omega_L t}$ and $\rho^{'}_{ii}$ = $\rho_{ii} $ represent the redefined matrix elements in the rotating frame. The total electric field amplitude is given by $G_{tot}= \sum_{j=1}^M G_j(t-t_j)$, and the detuning is defined as $\Delta_{01} = \omega_{01} - \omega_L$, where $\omega_L$ is the center frequency of the excitation pulse.
Each complex density matrix element is separated into real and imaginary parts.

As illustrated in Fig. \ref{fig:pulse}(b), the pulse sequence is defined with arrival times ($t_j$) for each pulse. The first pulse ($t_1$) can theoretically be set to zero; however, this would result in half of the pulse lying in the negative time domain. To avoid solving OBEs for negative time, is shifted to a positive time value, typically a few times the pulse duration ($\Delta t$).  For instance, in the \ref{Inhomo_Photon echo} case, with $\Delta t$ = 2.9 ps,  $t_1$ was set at 15 ps. Arrival times for subsequent pulses ($t_j$) are defined recursively as $t_j = t_1 + \tau_1 + .... + \tau_{j-1}$. The number of scanned or fixed delays can be adjusted based on the desired nonlinear signal and experimental dimensionality. For example, in rephasing FWM signals, $\tau_1$ and $\tau_3$ are scanned while $\tau_2$ remains fixed.

Numerical integration is performed using Python's odeint solver from Scipy library, with the time step set significantly smaller than the pulse duration to accurately resolve pulse overlap effects. For a 2.9 ps pulse, a time step of 0.25 ps was used. The integration time range is defined as $t_{total} = 2t_1 + \tau_1 + .....+\tau_{M-1}$ where the additional $t_1$ ensures no overlap between the last pulse and the detection time. Initial conditions are set such that the system starts in the ground state, with no population in the excited state or coherence terms.

Nested loops are implemented to scan parameters such as delays, electric field strengths, and phase cycling steps. The computations are parallelized using Python’s multiprocessing library to optimize efficiency and reduce runtime. This framework ensures accuracy in capturing system dynamics while maintaining computational feasibility.

\section{Optimized phase-cycling scheme}
\label{app:PCM_valid}
The phase-cycling method involves repeated measurements/calculations of the nonlinear signal for varying phase of the excitation pulses.
Consequently, an increase in the number of phase-cycling steps results in an increase in the total time for the simulation/experiment.
Thus, it is imperative to find the optimal phase-cycling scheme that uses minimum number of phase-cycling steps while successfully isolating the desired signal.
An essential step is to ensure that the chosen phase-cycling scheme works for the desired signal without aliasing \cite{Tan2008}.
We present an algorithm to validate the phase-cycling scheme defined by Eq. \eqref{eq:pcstep} for isolating the NWM signal under the phase matching condition defined by Eq. \eqref{eq:pm}.
\begin{enumerate}
  \item We list all possible nonlinear signals up to $N^{th}$ order light-matter interaction.
  We do this by finding all possible combinations of $\alpha'_j$ that satisfy the constraints defined by Eq. \eqref{eq:constraint} and
  \begin{equation}
    \sum_{j=1}^{M-1} \abs{\alpha'_{j}} + 1 \leq N.
    \label{eq:cond2}
  \end{equation}
The first condition ensures that the system necessarily ends in a population state after the last excitation pulse.
The second constraint ensures that we include only the contributions up to the $N^{th}$ order, including lower-order terms.
  
  \item We list signals originating due to aliasing of the required phase-matched signal, as defined by Eq. \eqref{eq:pm}.
  Specifically, we find all combinations of $\alpha_j + I P_j$, with integer $I \in [0,N)$.
  We have chosen the range of $I$ to include all relevant aliased signals.
  $I = 0$ corresponds to the phase-matching condition of the required signal.
  
  \item Finally, we check for validity of phase-cycling scheme by comparing the two lists.
  \begin{itemize}
    \item If \textbf{only one element is common} between the two lists (corresponding to the required signal), the phase-cycling scheme is \textbf{valid} and successfully isolates the NWM signal.
    \item If \textbf{more than one element is common}, it means that some aliased signals will mix with the desired signal, indicating that the phase-cycling scheme does not uniquely isolate the required signal. Thus, the phase cycling scheme is \textbf{invalid}.
    \end{itemize}
\end{enumerate}
We can iteratively increase the number of phase-cycling steps for each pulse $P_j \in [2,N+1]$ while using the above the algorithm to check whether it will work for the required signal.
The scheme with the minimum value of $N_{PC}$ that satisfies the above algorithm can be chosen as the optimum phase-cycling scheme.
This optimization process can be automated using a computer program to find the optimized phase-cycling scheme for any arbitrary NWM signal.

\section{Peak-amplitude calculation}
\label{app:noise}
We present the procedure used to calculate the FWM amplitude as function of the intensity of excitation pulses in Sec. \ref{saturation} .
One can use the peak of the 2D spectrum if the signal-to-noise ratio (SNR) and linewidth are independent of the intensity of the excitation pulses.
Both of these conditions are not satisfied for our measurements.
The SNR  improves and excitation-induced dephasing (EID) \cite{Li2006} leads to increase in the linewidth at higher intensities.
Thus, we need to estimate the amplitude while accounting for the above behaviour.

\begin{figure}[b]
\includegraphics{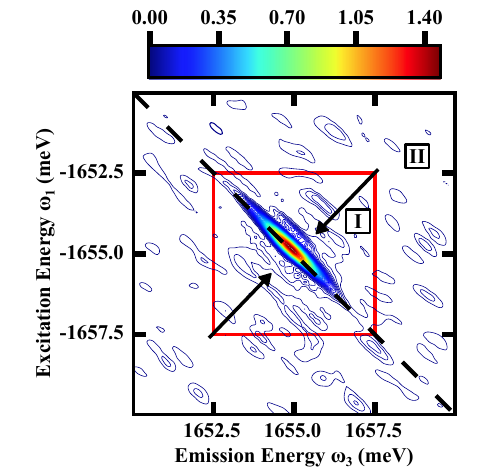}
\caption{2D intensity spectrum highlighting two rectangular sections: the inner section predominantly captures signal data points, while the outer section exclusively captures noise data points.
}
\label{fig:noise}
\end{figure}

The effect of the noise is more dominant at the lower intensity of the excitation pulses, which results in a higher estimation of the peak value.
Assuming white noise, we can consider the signal sitting on top of a white-noise floor in a 2D spectrum.
We remove the noise in the power spectrum, i.e. the square of the amplitude 2D spectrum.
We split the intensity 2D spectrum into two sections -- I and II, as shown in Fig. \ref{fig:noise} for the same data as in Fig. \ref{fig:2d_spec}(b).
The signal is confined in section I, whereas section II comprises of white noise.
We define the noise intensity as the average value of all the points in section II.

\begin{figure}[t]
\includegraphics{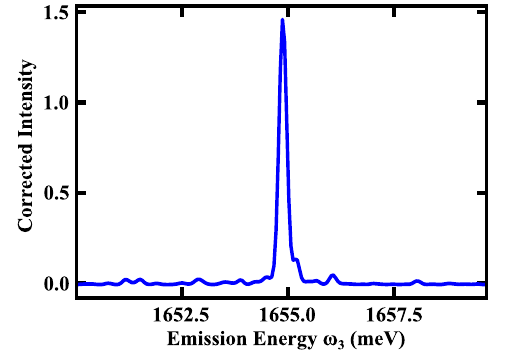}
\caption{Cross-diagonal slice through the data point with the maximum amplitude in the noise-corrected 2D intensity spectrum.}
\label{fig:noise2}
\end{figure}

EID results in broadening of the exciton peak as the excitation intensity is increased.
Consequently, the value of the peak amplitude will result in a lower estimate of the FWM amplitude.
We take a cross-diagonal slice across the peak of the noise-corrected spectrum, as shown in Fig. \ref{fig:noise2}.
All the values in this slice are summed to obtain a value proportional to total FWM signal intensity; integration along the frequency axis accounts for the effect of increase in the peak linewidth.
The integration of values outside the peak is nearly zero due to noise subtractions, which allows us to correctly estimate the signal strength.
A square root of the integrated intensity gives us the value of the FWM signal ampltitude.

The noise-correction and integration process is repeated for the 2D spectra obtained for various excitation intensities, which are shown in Fig. \ref{fig:pwr}.
We note that the precise frequency range used to define regions I and II does not significantly affect the resultant 2D spectrum, and illustrates the robustness of the aforementioned procedure.
We only need to ensure that we include a sufficiently large number of points in region II to reliably estimate the noise level.

\section{OBEs for diamond system}
\label{app:OBEs_diamond}
We present the OBEs for an exciton-biexciton diamond-type system, depicted in Fig. \ref{fig:bx}.
To ensure the cross-linear excitation scheme, we use a sequence of four pulses: the $1^{st}$ and $4^{th}$ are horizontally polarized, and the $2^{nd}$ and $3^{rd}$ are vertically polarized. The total electric field is given as:
\begin{equation}
E_{total}(t,r) = E^{H}_{1} + E^{V}_{2} + E^{V}_{3} + E^{H}_{4}
\end{equation}
where the electric fields $E^{j}_{i}(t,r)$ are defined similarly to those in Eq. \ref{eq:ej}, with the polarization (H or V) specified by the superscript $j$.

The OBEs for the density matrix elements of the diamond-type system are:
\begin{subequations}
\label{eq:obes_bx}
    \begin{align}
       {\dot{\rho}}_{HH} &= -\Gamma_1 \rho_{HH} +\Gamma_3 \rho_{BB} +\frac{i\mu_{GH}E_{H}}{\hbar} \Big(\rho_{GH} -\rho_{HG} \Big) \nonumber \\
      &\quad  +\frac{i\mu_{HB}E_{H}}{\hbar} \Big(\rho_{BH}-\rho_{HB} \Big), \\
        {\dot{\rho}}_{VV} &= -\Gamma_2 \rho_{VV}  +\Gamma_3 \rho_{BB} +\frac{i\mu_{GV}E_{V}}{\hbar}\Big(\rho_{GV}-\rho_{VG}\Big) \nonumber \\
       &\quad  +\frac{i\mu_{VB}E_{V}}{\hbar}\Big(\rho_{BV}-\rho_{VB}\Big) , \\
        {\dot{\rho}}_{BB} &= -2\Gamma_3 \rho_{BB} +\frac{i\mu_{HB}E_{H}}{\hbar}\Big(\rho_{HB}-\rho_{BH}\Big) \nonumber\\
       &\quad  +\frac{i\mu_{VB}E_{V}}{\hbar}\Big(\rho_{VB}-\rho_{BV}\Big) , \\
       {\dot{\rho}}_{GH} &= (i \omega_{HG} - \gamma_{GH})\rho_{GH} + \frac{i \mu_{GH} E_{H}}{\hbar}\big(\rho_{HH}-\rho_{GG}\big) \nonumber\\
       &\quad +\frac{i}{\hbar}\Big(\mu_{GV}E_V\rho_{VH} - \mu_{HB}E_H \rho_{GB}\Big) ,\\
       {\dot{\rho}}_{GV} &= (i \omega_{VG} - \gamma_{GV})\rho_{GV} + \frac{i \mu_{GV}E_V}{\hbar}\big(\rho_{VV}-\rho_{GG}\big) \nonumber\\
       &\quad +\frac{i}{\hbar}\Big(\mu_{GH}E_H\rho_{HV} - \mu_{BV}E_V \rho_{GB}\Big) ,\\
       {\dot{\rho}}_{HB} &= (i \omega_{BH} - \gamma_{HB})\rho_{HB} + \frac{i \mu_{HB} E_H}{\hbar}\big(\rho_{BB} - \rho_{HH}\big) \nonumber\\
       &\quad +\frac{i}{\hbar}\Big(\mu_{GH} E_H \rho_{GB}- \mu_{VB}E_V\rho_{HV}\Big) ,\\
       {\dot{\rho}}_{VB} &= (i \omega_{BV} - \gamma_{VB})\rho_{VB} + \frac{i\mu_{VB} E_V}{\hbar} \big(\rho_{BB} - \rho_{VV}\big) \nonumber\\
       &\quad + \frac{i E_H}{\hbar}\Big(\mu_{VG}\rho_{GB}- \mu_{HB}\rho_{VH}\big) ,\\ 
       {\dot{\rho}}_{HV} &= - \gamma_{HV}\rho_{HV} + \frac{i}{\hbar}\Big(\mu_{HG}E_H\rho_{GV} - \mu_{VB}E_V\rho_{HB}\Big)
       \nonumber\\
       &\quad +\frac{i}{\hbar}\Big(\mu_{HB}E_H\rho_{BV} - \mu_{VG}E_V \rho_{HG}\Big) ,\\
       {\dot{\rho}}_{GB} &= (i \omega_{BG}- \gamma_{GB})\rho_{GB} + \frac{i}{\hbar}E_H\Big(\mu_{GH}\rho_{HB} - \mu_{HB}\rho_{GH}\Big) \nonumber\\
       &\quad +\frac{i}{\hbar}E_V\Big(\mu_{GV}\rho_{VB} - \mu_{VB} \rho_{GV}\Big) ,
    \end{align}
\end{subequations}
where
\begin{itemize}
      \item $\rho_{ii}$ : Population in state $\ket{i}$
      \item $\rho_{ij}$ : Coherence between states $\ket{i}$ and $\ket{j}$
      \item $\mu_{ij}$ : Transition dipole moment for transition between states $\ket{i}$ and $\ket{j}$
      \item $\Gamma_{i}$ : Population decay rate for state $\ket{i}$ 
      \item $\gamma_{ij}$ : Coherence decay rate between states $\ket{i}$ and $\ket{j}$
      \item $\omega_{ij} = \omega_{i}-\omega_{j}$
\end{itemize}
with, $i\neq j =$ $G$, $H$, $V$, $B$, and $\omega_{GH} = \omega_{GV}$. 
Similar to Appendix \ref{app:Numerical_Solution}, we solve these OBEs after applying the RWA.

\begin{acknowledgments}
We thank Steven T. Cundiff, Daniel Gammon, and Allan S. Bracker for providing the interfacial quantum dot sample.
We thank Eric W. Martin for useful discussions regarding the 2DCS experiments.
The authors acknowledge support from the Science and Engineering Research Board (SERB), New Delhi under Project No. CRG/2023/003263.
K.K.M. and B.D. acknowledge the Ministry of Education, Government of India for support from the Prime Minister’s Research Fellows (PMRF) Scheme.
\end{acknowledgments}

\section*{Data Availability Statement}

The data that support the findings of this study are available from the corresponding author upon reasonable request.

\bibliography{Reference}

\end{document}